\documentclass[12pt,preprint]{aastex}
\usepackage{booktabs}

\newcommand{\lsim}{\raise0.3ex\hbox{$<$}\kern-0.75em{\lower0.65ex\hbox{$\sim$}}}
\newcommand{\gsim}{\raise0.3ex\hbox{$>$}\kern-0.75em{\lower0.65ex\hbox{$\sim$}}}
\newcommand{\propsim}{\raise0.3ex\hbox{$\propto$}\kern-0.75em{\lower0.65ex\hbox{$\sim$}}}

\usepackage{lineno}

\begin{document}
    
\title{Modelling of radio supernovae: Including the effects of inhomogeneities and radiative cooling}

\author{C.-I. Bj\"ornsson\altaffilmark{1}}
\altaffiltext{1}{Department of Astronomy, AlbaNova University Center, Stockholm University, SE--106~91 Stockholm, Sweden.}
\email{bjornsson@astro.su.se}

\begin{abstract}
The presence of inhomogeneities in a spatially unresolved source is often hard to establish. This limits the accuracy with which the source properties can be determined. It is shown how observed features not expected for a homogeneous model can be used to infer the properties of the inhomogeneities in radio supernovae. Furthermore, the observed consequences of radiative cooling can be seriously affected by inhomogeneities. It is shown that the deduced source properties are very sensitive to the observed value of the cooling frequency; even a lower limit is often useful to constrain its characteristics. It is argued that the main synchrotron emission region in SN\,2003L has a small volume filling factor, possibly as low as a few per cent. On the contrary, deviations from homogeneity are substantially smaller in SN\,2002ap. The observed properties of type Ib/c radio supernovae in general indicate the volume filling factor to remain rather constant with time for individual sources but those peaking later at radio frequencies have lower filling factors. The conditions in the main synchrotron component in both SN\,2003L and SN\,2002ap are consistent with equipartition of energy between relativistic electrons and magnetic fields.
\end{abstract}


\section{Introduction}
It is generally agreed that the radio emission observed from supernovae is due to synchrotron radiation. The relative simplicity of this emission process makes an analysis of the observations more straightforward than, for example, the optical emission. The standard model for the radio emission is equally simple - a homogeneous, spherically symmetric shell \citep{che82}. Its parameters can be constrained when a low frequency turn-over of the spectrum is observed, which is usually interpreted as either synchrotron self-absorption or free-free absorption. Another effect that can affect the observations is radiative cooling of the emitting electrons either due to the synchrotron emission itself or inverse Compton scattering of the radiation coming from the supernova. Sometimes, the result of the latter process can be observed as an emission component that falls mainly in the x-ray band. 

With a good radio spectrum, it is usually possible to distinguish between synchrotron self-absorption and free-free absorption. On the other hand, when x-ray emission is observed, it can be hard to determine what fraction, if any, is due to free-free emission. This makes type Ib/c supernovae especially interesting, since the wind velocity from their progenitor stars is thought to be much higher than that appropriate for type II supernovae. On average, the mass-loss rates of the progenitor stars are not likely to differ that much between the two types of supernovae. Hence, the density of the circumstellar medium, with which the ejecta from  type Ib/c supernova interacts, is much lower than that corresponding to type II supernovae. It is thought that the strength of the magnetic field scales with the thermal energy density of the shocked gas. For the radio emission region, this leads to a lower density in type Ib/c supernovae as compared to type II. Since free-free emission scales as density square, while the inverse Compton scattered emission is expected to scale, roughly, linearly with density, the relative contribution to the x-ray emission from free-free emission should be substantially lower in type Ib/c supernovae than type II; in fact, as discussed in \cite{bjo13}, observations suggest that inverse Compton scattering dominates the emission in the x-ray band. Hence, detection of x-ray emission from type Ib/c supernovae is likely to provide a rather robust constraint of the properties of the synchrotron emission region.

Although type Ib/c is the class of supernovae in which inverse Compton scattering is likely to be most dominant, there are indications of a substantial contribution to the x-ray emission also in other types of supernovae; for example, in the type IIb supernova 2011dh \citep{sod12,kra12,hor13}. Even in type IIPs, in which the density of  the circumstellar medium is thought to be much higher than in type Ib/c, it has been argued that inverse Compton scattering gives an important contribution to the x-ray emission; for example, SN\,2004dj \citep{cha12,nay18} and SN\,2016X \citep{r-c22}.

Radiative cooling by itself causes a relative distinct break in the energy distribution of the emitting electrons; however, the presence of adiabatic cooling smears out the transition from the low to the high energy electrons \citep{bjo22}. This transition from adiabatic cooling in the low frequency range to radiative cooling at high frequencies introduces a curvature in the observed optically thin synchrotron spectrum. Since radiative cooling affects the synchrotron emission over a large frequency range, one would expect its effects to be observed in many sources. However, the cooling frequency is hard to determine observationally; often, it is estimated from visual inspection of the spectra and/or light curves. There are only a few claimed detections of its effects on the observed radio emission; for example, SN\,2020oi \citep{hor20},  SN\,2012aw \citep{yad14} and SN\,2013df \citep{kam16}. As discussed in \cite{bjo22}, this is likely due to the slow spectral transition together with the rather narrow spectral range usually available, which makes it hard to measure the spectral curvature. In order to reliably determine a value for the cooling frequency (or a lower limit), a curved spectrum needs to be used in the fitting process directly. 

One aspect of the standard model that affects the deduced source properties is the assumption of homogeneity; for example, the observed x-ray emission often implies a very high value for the ratio of energy densities in relativistic electrons and magnetic fields in the synchrotron emission region. However, it could also indicate an inhomogeneous source structure \citep{bjo13}, i.e., that the emitting volume for the inverse Compton scattered radiation is larger than for the synchrotron emission. It is important to be able to distinguish between these two alternatives, since the implications for the physical processes responsible for the acceleration of electrons as well as the amplification of the magnetic field are quite different in the two scenarios. Therefore, it is essential to recognise observed properties, which deviate from those  expected in the standard model, since they are a good starting point for establishing the presence of inhomogeneities. 

The most direct indication for the presence of inhomogeneities is flat-topped spectra/light curves. A discussion was given in \cite{b/k17} of the relation between observed spectral deviations and the qualitative characteristics that could be deduced for the inhomogeneities. However, one should note that, even if the standard model gives a good fit to observations, it is not possible to conclude that it also gives a fair description of the source properties; an example of such a case is SN\,1993J. 

\cite{f/b98} showed that both its instantaneous radio spectra and light curves after a few hundred days were well described by the standard model. The proximity of SN\,1993J made it possible to determine its outer radius independently via VLBI-observations. Assuming a homogeneous source, the measured brightness temperature was lower than expected for energy equipartition beteen relativistic electrons and magnetic fields, which implied a source that was strongly magnetically dominated. However, as discussed in \cite{bjo15}, there are several indications that such an interpretation is not correct. Instead, it was shown that an inhomogeneous model together with equipartition between relativistic electrons and magnetic fields could give a consistent description of the observations. The difference to the supernovae with flat-topped spectra discussed in \cite{b/k17} would then be that radio emission in SN\,1993J comes from regions with a smaller range of values for the magnetic field (alternatively, optical depths). It may also be noted that VLBI-observations showed intensity fluctuations \citep{bie03} in the spatially resolved source. However, they were not large enough to account for the low brightness temperature. Hence, the inhomogeneities in SN\,1993J were dominated by fluctuations on scales smaller than that resolved by VLBI. In fact, it was argued that the structure of SN\,1993J was quite similar to the spatially resolved radio emission observed in the supernova remnant Cassiopeia\,A, which was also a type IIb supernova.

One of the aims of the present paper is to provide a quantitative description of the inhomogeneities. To this end, in section \ref{sect2b}, a modified version of the standard model is developed, from which the source structure can be deduced from observations. This formulation is guided by simplicity with the hope of being user friendly. Another aim is to emphasize the importance of radiative cooling. It is pointed out in section \ref{sect2c} that even a lower limit to the cooling frequency can often give useful constraints on the source properties. In addition, it is shown how the effects of inhomogeneities and radiative cooling are intertwined. In order to illustrate these results, a reanalysis of the observations of SN\,2003L and SN\,2002ap is done in section \ref{sect3}, with special attention to the role played by radiative cooling. A discussion of the results follows in section \ref{sect4}, where, in particular, the implications for the partition of energy between relativistic electrons and magnetic field are considered. The conclusions of the paper are summerized in section \ref{sect5}.  Numerical results are mostly given using cgs-units. When this is the case, the units are not written out explicitly.

\section{Synchrotron radiation and inverse Compton scattering}\label{sect2}
In order to study the effects of inhomogeneities, it is useful to start with the standard homogeneous model and then introduce modifications, which can account for various physically relevant deviations from homogeneity. The modifications will be limited to those that have the potential to be constrained by radio and x-ray observations.

\subsection{A homogeneous, spherically symmetric source}\label{sect2a}
In addition to homogeneity and sphericity, the standard synchrotron model assumes the radiating electrons to have a distribution of Lorentz factors ($\gamma$) according to  $n(\gamma) = K_{\rm o} \gamma^{-{\rm p}}$ for $\gamma > \gamma_{\rm min}$ and $\rm {p\,>\,2}$. The deduced magnetic field strength can then be expressed as
\begin{equation}
	B = 1.0\frac{\nu_{{\rm abs},10}}{\left(y^2 F_{\nu_{\rm abs},27}\right)^{2/19}},
	\label{eq1}
\end{equation}
and the source radius is
\begin{equation}
	R_{16}= 0.55\frac{F_{\nu_{\rm abs},27}^{9/19}}{y^{1/19}\nu_{\rm abs,10}},
	\label{eq2}
\end{equation}
where $R_{16} \equiv R/10^{16}$ and p\,=\,3 has been assumed. Since the optically thin spectral index $\alpha = (p-1)/2$ is close to 1 for most type Ib/c supernovae \citep{c/f06}, for convenience, this value of p will be used throughout the paper. The expressions for arbitrary p can be found in \cite{bjo21}. Here,  $\nu_{\rm abs}$ is the frequency where the spectral flux peaks and $F_{\nu_{\rm abs}}$ is the corresponding spectral flux. Furthermore, $\nu_{{\rm abs},10} \equiv \nu_{\rm abs}/10^{10}$ and $F_{\nu_{\rm abs},27} \equiv F_{\nu_{\rm abs}}/10^{27}$. 

The expression for $y$ is given by
\begin{equation} 
	y = \gamma_{\rm min}\frac{U_{\rm rel}}{U_{\rm B}}\frac{R_{||}}{R},
	\label{eq3}
\end{equation}
where $R_{\rm ||}$ is the average line of sight extension of the source, $U_{\rm rel}$ and $U_{\rm B}$ are the energy densities of the relativistic electrons and the magnetic field, respectively. The value of  $R_{\rm ||}$ is twice the shell thickness and, for example,  $R_{\rm ||}/R = 1/2$ for a shell resulting from a strong forward shock. The definition of $y$ in equation (\ref{eq3}) is the combination of three different internal characteristics of the source; namely, the injection of electrons into the acceleration process ($\gamma_{\rm min}$), the partition of energies between electrons and magnetic fields ($U_{\rm rel}/U_{\rm B}$) and the relative thickness of the source ($R_{||}/R$). The values of the first two parameters are determined by processes that are not so well understood physically. Furthermore, one may note that it is the parameter $y$, which can be determined directly from observations. Hence, for example, a value for $U_{\rm rel}/U_{\rm B}$ can be obtained only when the values of both $\gamma_{\rm min}$ and  $R_{\rm ||}/R $ are known; in particular the latter value is quite sensitive to the presence of inhomogeneities (see below).

It order to find the value for $y$, additional observations are needed. In supernovae, the optical emission ($L_{\rm bol}$) can be inverse Compton scattered by the same electrons giving rise to the radio emission. When this radiation ($L_{\rm x}$) is observed, the value of $y$ is obtained from
\begin{equation}
	y = 0.23\,F_{\nu_{\rm abs, 27}}^{7/5}\left(\frac{L_{\rm x}}{L_{\rm radio} L_{\rm bol, 42}}\right)^{19/10},
	\label{eq4}
\end{equation}
where $L_{\rm bol, 42} \equiv L_{\rm bol}/10^{42}$ \citep{bjo22}. The luminosities are here defined as $L \equiv \nu F_{\nu}$, where $F_{\nu}$ is the optically thin spectral flux at a frequency $\nu$.  This then leads to a closure relation for the three independent parameters $B$, $R$ and $y$ in the standard model. It should be noted, though, that it is only the parameter $y$, which is directly related to observed quantities (equation (\ref{eq4})). The values of $B$ and $R$ are only indirectly so, since they depend on $y$ (equations (\ref{eq1}) and (\ref{eq2})). This distinction will be important in the discussion below of the parameterization of the inhomogeneities.

The optically thin emission extrapolated  to $\nu_{\rm abs}$ is a factor $\tau_{\rm hom}/(1-\exp(-\tau_{\rm hom}))$ larger than $F_{\nu_{\rm abs}}$, where $\tau_{\rm hom}$ is the synchrotron optical depth at the spectral peak. Since  $\tau_{\rm hom} = 0.64$ \citep{bjo21}, $L_{\rm radio} =1.4\,\nu_{\rm abs}F_{\nu_{\rm abs}}$ and equations (\ref{eq1}), (\ref{eq2}) and (\ref{eq4}) can be written
\begin{equation}
	B = 1.6 \nu_{\rm abs,10}^{7/5}\left(\frac{L_{\rm bol,42}}{L_{\rm x,37}}\right)^{2/5},
	\label{eq1a}
\end{equation}
\begin{equation}
	R_{16} = 0.61\frac{F_{\nu_{\rm abs,27}}^{1/2}}{\nu_{\rm abs,10}^{9/10}} \left(\frac{L_{\rm bol,42}}{L_{\rm x,37}}\right)^{1/10},
	\label{eq2a}
\end{equation}
and
\begin{equation}
	y = 0.13 \frac{1}{F_{\nu_{\rm abs,27}}^{1/2}\nu_{\rm abs,10}^{19/10}}\left(\frac{L_{\rm x,37}}{L_{\rm bol,42}}\right)^{19/10},
	\label{eq4a}
\end{equation}
where $L_{\rm x,37} \equiv L_{\rm x}/10^{37}$.

\subsection{Inhomogeneities}\label{sect2b}
One example of accounting for deviations from the standard model was given by \cite{sod05}. In order to fit the observations of the type Ib/c supernova 2003L, they introduced a parameter $\xi$, which modified the synchrotron spectral emissivity according to 
\begin{equation}
	f_{\xi}(\nu) \propto \frac{\nu^{5/2}}{B^{1/2}}\left[1-\exp(-\tau^{\xi}(\nu))\right]^{1/\xi},
	\label{eq5}
\end{equation}
where $\tau(\nu)$ is the synchrotron optical depth. Hence, $\xi = 1$ corresponds to the true synchrotron emissivity. Such a modification flattens ($\xi <  1$) the spectral emissivity around its peak, while conforming to the true synchrotron emissivity for $\tau \gg 1$ and $\tau \ll 1$. It should be noted that $\tau(\nu)$ was assumed to have the value given by a homogeneous synchrotron source so that values of $B$ and $R$ could be derived from expressions corresponding to equations (\ref{eq1}) and (\ref{eq2}). 

One may also note that equation (\ref{eq5}) amounts to a modification of the underlying physics rather than the physical model. The meaning of the values of $B$ and $R$, thus derived, is not clear, since they are based on an artificial emissivity. However, as discussed in \cite{b/k17}, $\xi$ is a useful parameter, since $f_{\xi}(\nu)$ can be given a physical interpretation. It can be understood as the averaged/smeared out emission from an inhomogeneous source. The flattening around the spectral peak can then be described as an overlap of homogeneous sources, each of which with a different value for $\nu_{\rm abs}$. This implies that the volume filling factor of these sub-components is smaller than unity and it is convenient to write it as $\phi_{\rm fil} =\phi_{\rm cov}\times \phi_{los}$, where $\phi_{\rm cov}$ is the covering factor and $\phi_{\rm los}$ is the reduced line of sight extension of a sub-component. Also, for convenience, in the discussion below, the subscript will be dropped on $\nu_{\rm abs}$ in the flattened spectral region so that, for example, $F_{\nu_{\rm abs}} \equiv F(\nu)$.

The aim of the discussion in \cite{b/k17} was to use the spectral flux variations in the flattened part of the spectrum ($\nu_{\rm min} < \nu < \nu_{\rm max}$) to constrain the characteristics of the inhomogeneities. This made possible a qualitative description of the source properties. In the present paper, instead, the focus is on quantitative results together with inclusion of constraints provided by an observed x-ray emission. The assumption of overlapping homogeneous sub-components is, of course, an approximation to a likely continuous distribution of magnetic field strengths and densities of relativistic electrons. Since the local synchrotron spectral emissivity has a finite width ($\Delta \nu \sim \nu$), the properties of a sub-component peaking at a frequency $\nu$ correspond to those averaged over that spectral range; for example, $F(\nu)$ gives the emission averaged value of $B$ for the parts of the source contributing to the flux at frequency $\nu$. Particular attention will be given to the boundaries (i.e., $\nu_{\rm min}$ and $\nu_{\rm max}$), since they are important for determining the expansion velocity of the forward shock and the possible origin of the x-ray emission. It will be assumed that they are distinct, i.e., for $\nu < \nu_{\rm min}$, $F_{\nu} \propto \nu^{5/2}$ and $F_{\nu} \propto \nu^{-1}$ for $\nu > \nu_{\rm max}$. 

The above characterization of the inhomogeneities differs from that used in \cite{b/k17}. The physically relevant property of the source is the structure of the magnetic field. Hence, the distribution of B-values was taken as the starting point for the qualitative discussion in that paper; for example, filling and covering factors were given in terms of magnetic field values. However, the value of the magnetic field cannot be directly observed but has to be deduced. Here, the observable  $\nu_{\rm abs}$ is used instead of the parameter B, since it gives a more model-independent description of the inhomogeneities. In addition, one may then regard the inhomogeneities as a distribution of homogeneous sub-components, which, in turn, facilitates a quantitative estimate of the properties of the inhomogeneities. However, it may be noticed that the same degeneracy discussed in \cite{b/k17} also appears here; namely, that a given value of $\nu_{\rm abs}$ corresponds to a combination of B-values and column densities of relativistic electrons. These cannot be separated individually unless the radio source is spatially resolved (see also below).

Each of the homogeneous sub-components is then described by equations (\ref{eq1}) and (\ref{eq2}). The modifications introduced by inhomogeneities affect the parameter $y$ only. The definition of $y$ in equation (\ref{eq3}) remains valid, except that the value of $R_{||}(\nu)/R(\nu)$ may now exceed 1/2. Furthermore, the expression for $y$ in equation (\ref{eq4}), which relates its value to observed quantities, needs to be modified when x-ray emission is observed. In the standard model, both $L_{\rm radio}$ and $L_{\rm x}$ are proportional to $e_{\rm rel} \equiv \gamma_{\rm min}U_{\rm rel} R_{||}$. Since all the relativistic electrons in the source contribute to $L_{\rm x}$, while $L_{\rm radio}(\nu)$ is determined only by those in the sub-component, $L_{\rm radio}/L_{\rm x}$ should be substituted by $(L_{\rm radio}(\nu)/L_{\rm x})(e_{\rm rel,ave}/e_{\rm rel}(\nu))$, where $e_{\rm rel, ave}$ is the surface averaged value of $e_{\rm rel}(\nu)$. This gives
\begin{equation}
	B(\nu) = 1.6\,\nu_{10}^{7/5}\left(\frac{L_{\rm bol,42}}{L_{\rm x,37}}\right)^{2/5}\left(\frac{e_{\rm rel}(\nu)}{e_{\rm rel,ave}}\right)^{-2/5},
	\label{eq6}
\end{equation}
\begin{equation}
	R_{16}(\nu) = 0.61\,\frac{F(\nu)^{1/2}_{27}}{\nu_{10}^{9/10}} \left(\frac{L_{\rm bol,42}}{L_{\rm x,37}}\right)^{1/10}\left(\frac{e_{\rm rel}(\nu)}{e_{\rm rel,ave}}
	\right)^{-1/10},
	\label{eq7}
\end{equation}
and
\begin{equation}
	y(\nu) = 0.13\,\frac{1}{F(\nu)_{{27}}^{1/2}\nu_{10}^{19/10}}\left(\frac{L_{\rm x,37}}{L_{\rm bol,42}}\right)^{19/10}\left(\frac{e_{\rm rel}(\nu)}{e_{\rm rel,ave}}\right)^{19/10}.
	\label{eq8}
\end{equation}
Although equations (\ref{eq6}), (\ref{eq7}), and (\ref{eq8}) are just the standard equations rewritten for an inhomogeneous source, they have the additional advantage that the impact of the observed radio properties (i.e., $F(\nu)$ and $\nu$) are explicitly separated from those of the x-rays (i.e., $L_{\rm x}$). It is seen that the effects of the inhomogeneities are described by $e_{\rm rel}(\nu)$, which is, basically, the energy column density of relativistic electrons (i.e., the optical depth to Thomson scattering). 

Let $R_{\rm o}$ denote the actual radius of the source. The covering factor can then be written $\phi_{\rm cov}(\nu) \approx [R(\nu)/R_{\rm o}]^2$ and $\phi_{\rm los}(\nu) = 2 R_{||}(\nu)/R_{\rm o}$.  In order to simplify the discussion, it will be assumed that $R(\nu_{\rm min}) \approx R_{\rm o}$. This implies that the value of $R(\nu_{\rm min})$ gives a lower limit to the expansion velocity of the forward shock. Furthermore, for convenience, $R_{\rm min} \equiv R(\nu_{\rm min})$ and $R_{\rm max} \equiv R(\nu_{\rm max})$ will be used (with a similar notation for the other frequency dependent variables). Hence, from equation (\ref{eq7})
\begin{equation}
	\phi_{\rm cov}(\nu) = \frac{F(\nu)}{F_{\rm min}}\left[\frac{\nu}{\nu_{\rm min}}\right]^{-9/5}\left[\frac{e_{\rm rel}(\nu)}{e_{\rm rel,min}}\right]^{-1/5}.
	\label{eq9}
\end{equation}

With a good spectrum, $\nu_{\rm min}$ and $\nu_{\rm max}$ can be estimated. In many cases, the light curves are better sampled than the spectra and it is common to model light curves with an assumed spectrum, including its temporal evolution. It should be noted that even for a homogeneous source, at a given time, the spectral peak frequency does not in general coincide with the frequency for which the light curve peaks. Hence, equations (\ref{eq1}) and (\ref{eq2}) should not be used to derive values for $B$ and $R$ using light curves and their corresponding peak fluxes. However, as was shown in \cite{bjo22}, there is one situation when these two frequencies coincide, namely, when $F_{\nu_{\rm abs}}$ is constant with time.

\subsection{Radiative cooling}\label{sect2c}
The standard homogeneous synchrotron source model has three free parameters; namely, $B, R$ and $y$. As discussed above, they can all be determined, if the inverse Compton scattered radiation can be observed in addition to the self-absorbed synchrotron radiation. Radiative cooling is another physical process, which can be used to constrain the source structure. Together, such observations allow to investigate the basic assumption of source homogeneity. 

The radiative cooling time is given by $t_{\rm cool} = 1/a\gamma$ with $a = 4\sigma_{\rm T}U/(3mc)$, where $U$ is the total energy density of photons ($U_{\rm ph}$) and magnetic fields ($U_{\rm B}$), and $\sigma_{\rm T}$ is the Thomson cross-section. It is common to define the cooling frequency ($\nu_{\rm cool}$) as the frequency for which $t_{\rm cool} = t_{\rm ad}$, where $t_{\rm ad}$ is the adiabatic cooling time. Hence, $\nu_{\rm cool} =1.6\nu_{\rm B}/(at_{\rm ad})^2$, where $\nu_{\rm B}$ is the cyclotron frequency and the numerical factor is appropriate for optically thin synchrotron radiation and p\,=\,3 \citep{bjo22}. Since $U_{\rm ph}/U_{\rm B} = 2L_{\rm bol}/c(BR)^2$, one finds from equations (\ref{eq1}) and (\ref{eq2})
\begin{equation}
	\nu_{{\rm cool},10} = 3.5\times10^2 \frac{F^{6/19}_{{\nu_{\rm abs}},27}\,y^{12/19}}{\nu^3_{{\rm abs},10}\,t_{{\rm ad},10}^2}\frac{1}{\left[1+2.2\,L_{{\rm
	bol},42}\,y^{10/19}/
	F^{14/19}_{{\nu_{\rm abs}},27}\right]^2},
	\label{eq10}
\end{equation}
where $\nu_{{\rm cool},10} = \nu_{\rm cool}/10^{10}$ and $t_{{\rm ad},10} = t_{\rm ad}/10\,$days. One should note that the value for $\nu_{\rm cool}$ does not involve the values of $B$ and $R$ explicitly only the value for $y$. Hence, for a self-absorbed synchrotron source, $y$ is the central parameter, since its value determines not only $\nu_{\rm cool}$ but also $B$ and $R$ as well as the combination of source properties $\gamma_{\rm min}(U_{\rm rel}/U_{\rm B})(R_{||}/R)$ (section\,\ref{sect2a}). 

The cooling time is usually defined as ${\rm d}\gamma/{\rm d}t \equiv -\gamma/t_{\rm cool}$. For adiabatic cooling, $\gamma \propto n^{1/3}$, where $n$ is the density of particles. Behind a propagating shock, $n \propto R_{\rm o}^{-2}$. Together with the assumption $R_{\rm o} \propto t^{\rm m}$, this leads to $t_{\rm ad} = 3t/2{\rm m}$, where $t$ is the time since the start of the expansion. Due to the possibility of a reduced line of sight extension in an inhomogeneous source, the effective time for cooling is $\phi_{\rm los} t$. Hence, the radiative cooling time should be compared to an effective adiabatic cooling time given by  $t_{\rm ad}^{\rm eff} = 3\phi_{\rm los} t/2{\rm m}$. By substituting $t_{\rm ad}^{\rm eff}$ for $t_{\rm ad}$ in equation (\ref{eq10}), it is also valid for the homogeneous subcomponents in an inhomogeneous source. In practice, this is limited to the sub-component with $\nu_{\rm abs} = \nu_{\rm max}$, since this component dominates both the cooling and the optically thin synchrotron emission. 

\begin{figure}
\plotone{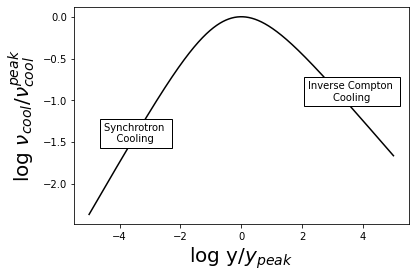}
\caption{The variation of the cooling frequency ($\nu_{\rm cool}$) with the parameter $y$ (equation (\ref{eq3})). The normalization parameters $\nu_{\rm cool}^{\rm peak}$ and $y_{\rm peak}$ are defined in equations (\ref{eq10a}) and (\ref{eq10b}). The range of parameter-values for which cooling is dominated by synchrotron radiation and inverse Compton scattering, respectively, are indicated.
\label{fig1}}
\end{figure}

The values of both $B$ and $R$ decrease for larger $y$ (see equations (\ref{eq1}) and (\ref{eq2})). Since $U_{\rm ph}\propto R^{-2}$, this implies that synchrotron cooling dominates for small values of $y$, while cooling due to inverse Compton scattering takes over for large values. Hence, there is a maximum value for $\nu_{\rm cool}$, which can be calculated from equation (\ref{eq10}) as
\begin{equation}
	\nu_{\rm cool,10}^{\rm peak} = 1.6\times 10\,\frac{m^2}{\phi_{\rm los}^2}\frac{1}{\nu^3_{{\rm abs},10}\,t_{10}^2}\left(\frac{F_{\nu_{\rm abs,27}}}{L_{\rm bol,42}}\right)^{6/5},
	\label{eq10a}
\end{equation}
where $t_{10} \equiv t/10$\,days. The corresponding value of $y$ is given by
\begin{equation}
	y_{\rm peak}= 0.49\frac{F_{\nu_{\rm abs,27}}^{7/5}}{L_{\rm bol,42}^{19/10}}.
	 \label{eq10b}
\end{equation}
The expression for the cooling frequency in equation (\ref{eq10}) can then be rewritten
\begin{equation}
	\hat{\nu} = \frac{25}{4}\frac{\hat{y}^{12/19}}{\left(1+ \frac{3}{2}\hat{y}^{10/19}\right)^2},
	\label{eq10c}
\end{equation}
where $\hat{\nu} \equiv \nu_{\rm cool}/\nu_{\rm cool}^{\rm peak}$ and $\hat{y} \equiv y/y_{\rm peak}$. This relation is shown in figure\,\ref{fig1}, and, together with equations (\ref{eq10a}) and(\ref{eq10b}), it summarizes the combined effects of cooling and inhomogeneities. One may note that these effects are intertwined, i.e., when only the cooling frequency is observed, the solution is degenerate, since the value of $y$ depends on $\nu_{\rm cool}\,\phi_{\rm los}^2$.

It is seen that $\nu_{\rm cool} = \nu_{\rm cool}^{\rm peak}$ corresponds to $U_{\rm ph}/U_{\rm B} = 3/2$. Furthermore, an observed value of $\nu_{\rm cool} >  \nu_{\rm cool}^{\rm peak}(\phi_{\rm los}=1)$ implies that the standard homogeneous model is not valid and, instead, can be used to set an upper limit to $\phi_{\rm los}$. This conclusion is independent of $L_{\rm x}$. As is seen in figure\,\ref{fig1}, the two solutions to equation (\ref{eq10c}) correspond to synchrotron cooling ($y < y_{\rm peak}$) and inverse Compton cooling ($y > y_{\rm peak}$). Another feature to note from figure\,\ref{fig1} is the $\hat{\nu}(\hat{y})$ is quite flat-topped around its peak. This implies that even small variations of $\nu_{\rm cool}$ around its peak value leads to rather larger changes in the ratio of the two values for $y$. Hence, unless $\nu_{\rm cool}$ is close to its peak value, the choice between synchrotron and inverse Compton cooling is expected to lead to very different values for $B$, $R$ and/or $\phi_{\rm los}$; for example, $\hat{\nu} \approx 1/3$ leads to a ratio between the two $y$-values of $\approx 10^3$. With the assumption that the inhomogeneities are the same in the two cases, this causes a difference in the deduced $B$-values by a factor $\approx 5$ and $R$-values by $\approx 1.5$ (equations (\ref{eq1}) and (\ref{eq2})).

If $L_{\rm x}$ is also measured, the main cooling process is determined directly for a homogeneous source, since then $U_{\rm ph}/U_{\rm B} = L_{\rm x}/L_{\rm radio}$. The appropriate value of $y$ can be calculated from equation (\ref{eq10}) and compared to the corresponding value obtained from equation (\ref{eq4a}). The homogeneous model requires these two values to be equal; if not, an inhomogeneous source is indicated. Hence, for an inhomogeneous source, observation of $L_{\rm x}$ does not help to discriminate between synchrotron and inverse Compton cooling. Instead, it provides constraints on the column density of relativistic electrons in the two scenarios (see equations (\ref{eq6}) - (\ref{eq8})).  In addition to the two solutions appearing for an inhomogeneous source, one may also note that there are two new parameters ($\phi_{\rm los}$ and $e_{\rm rel}/e_{\rm rel,ave }$) but only one new observable ($\nu_{\rm cool}$). Hence, further observations and/or constraints are needed to find the parameter-values. As will be discussed in section \ref{sect4b}, the incorporation of all the available information often necessitates a more complex analysis than that appropriate for the standard model; for example, a forward modelling approach. However, in order to illustrate how a few physically motivated assumptions can be used to deduce a reasonable range of parameter-values, the observations of SN\,2003L and SN\,2002ap are discussed in section \ref{sect3}. 

This expression for $t_{\rm ad}^{\rm eff}$ assumes that radiative cooling occurs in the radio emission region only. This is true for synchrotron cooling but care needs to be taken when the cooling is dominated by inverse Compton scattering, since, in this case, the mode of injecting electrons affects the result. When electrons are injected uniformly over the source (e.g., turbulent acceleration), the situation corresponds, roughly, to that of synchrotron cooling. However, the position of the synchrotron emission region is important for electrons injected at the forward shock. For an emission region close to the forward shock, the result is, again, similar to the synchrotron cooling case. On the other hand, the electrons will suffer radiative cooling prior to entering the synchrotron emission region, if it is located some distance away from the forward shock. This corresponds to a situation in which the energy distribution of injected electrons is curved rather than having a power-law appearance. 

In the case when the inferred value of $\nu_{\rm cool}$ is larger than the maximum possible for a homogeneous source, an upper limited can be obtained for $\phi_{\rm los,max}$. However, further restrictions on the values of $y_{\rm max}$ and $\phi_{\rm los,max}$ are harder to come by. It is seen from equation (\ref{eq8}) that a comparison with the relevant value of $y_{\rm max}$ obtained from $L_{\rm x}$ also involves the density of relativistic electrons, i.e., $e_{\rm rel,max}/e_{\rm rel,ave } = \delta_{\rm max} \phi_{\rm los,max}$, where $\delta_{\rm max} \equiv U_{\rm rel,max}/U_{\rm rel,ave}$ (where, for simplicity, $\gamma_{\rm min}$ has been assumed to be constant over the source). Hence, assumptions need to be made regarding the spatial distribution of relativistic electrons.

The reduction of the column density of relativistic electrons due to radiative cooling affects the emitted radiation in two ways. If the cooling frequency falls in the optically thin part of the spectrum ($\nu_{\rm cool}\gg \nu_{\rm abs}$), the local spectral index will increase with frequency as the dominant cooling process changes from adiabatic to radiative, resulting in a curved spectrum. When $\nu_{\rm cool} \sim \nu_{\rm abs}$, the main effect is instead decreasing values for both $F_{\nu_{\rm abs}}$ and $\nu_{\rm abs}$. 

As already mentioned, the value of $\nu_{\rm cool}$ is usually hard to determine. The most direct way comes about when the cooling frequency falls in the optically thin part of the spectrum, since, in principle, the spectral curvature can be used to derive its value. The spectral curvature is given by $\Delta \alpha = \Delta p/2$, where $\Delta p = 1/[1 + (\nu_{\rm cool}/\nu)^{1/2}] $ is the steepening of the electron energy distribution due to cooling \citep{bjo22}. It is seen that the transition from the adiabatic to the radiative part of the spectrum is relatively slow. This is due to adiabatic cooling, which smears out the distinct spectral break expected from radiative cooling alone. Therefore, in order to measure the spectral curvature, good observations over a wide spectral range are normally needed. This is rarely the case; in addition, the best observations are usually done around flux maximum, which, for supernovae, corresponds to frequencies close to $\nu_{\rm abs}$. This makes it hard to separate the spectral curvature caused by radiative cooling from that due to optical depth effects.

An alternative is to use an indirect way to estimate  $\nu_{\rm cool}$ by varying the assumptions regarding the observed emission in the absence of cooling (i.e., forward modelling). A comparison to the actual observations can then give a likely value for $\nu_{\rm cool}$ and, hence, the source properties \citep[see][for a more detailed discussion]{bjo22}. Assuming a value for the spectral index (e.g., $\alpha = 1$), the spectral curvature is given directly by $\Delta \alpha = \alpha_{\rm obs} - \alpha$, where $\alpha_{\rm obs}$ is the  observed optically thin spectral index obtained in a fitting procedure using the standard model without radiative cooling. Likewise, assumptions on the time variation of $F_{\nu_{\rm abs}}$ (e.g., $F_{\nu_{\rm abs}} =$ constant) or $\nu_{\rm abs}$ can be used to calculate changes of the column density of electrons due to radiative cooling. The reduction of the column density is given by $1 + (\nu/\nu_{\rm cool})^{1/2}$. Hence, for a given source, $F_{\nu_{\rm abs}} \propto y^{5/7} \propto [1 + (\nu_{\rm abs}/\nu_{\rm cool})^{1/2}]^{-5/7}$.

\section{Observations of SN\,2003L and SN\,2002ap}\label{sect3}
The number of supernovae of type Ib/c with both radio and x-ray observations are rather limited. The reason for choosing SN\,2003L and SN\,2002ap for a more detailed study is that the former has the most extensive radio observations both spectrally and temporally, while the latter show indications of radiative cooling in the radio spectra \citep{b/f04}. In addition, SN\,2002ap is the earliest one observed with the radio light curves peaking within 10 days after the supernova explosion, which makes it possible to consider structural differences between supernovae peaking at radio frequencies at different times.

\subsection{SN\,2003L}\label{sect3a}
\cite{sod05} found the light curves of SN\,2003L to be well described by a constant value $\xi = 0.5$. As shown in the Appendix, this gives for the inhomogeneities in SN\,2003L 
\begin{equation}
	\phi_{\rm cov,max} = 0.25 \left[\frac{e_{\rm rel,max}}{e_{\rm rel,min}}\right]^{-1/5},
	\label{eq14}
\end{equation}
\begin{equation}
	\frac{B_{\rm max}}{B_{\rm min}} = 2.9 \left[\frac{e_{\rm rel,max}}{e_{\rm rel,min}}\right]^{-2/5},
	\label{eq15}
\end{equation}
and
\begin{equation}
	\frac{y_{\rm max}}{y_{\rm min}} = 0.24\left[\frac{e_{\rm rel,max}}{e_{\rm rel,min}}\right]^{19/10}.
	\label{eq16}
\end{equation}

Since the observations of SN\,2003L are consistent with a constant value for $\xi$, it is likely that the characteristics of the inhomogeneities remain fairly stable. The various light curves peak at roughly the same spectral flux, which suggests that the same is true for the sub-components. As discussed in section\,\ref{sect2b}, this causes the light curve for a given frequency to peak at the same time as that frequency corresponds to the spectral peak; in particular, the peak flux of a light curve for a frequency $\nu$ corresponds to the spectral peak flux of the sub-component with $\nu = \nu_{\rm p}$ , where $\nu_{\rm p}$ is the peak frequency of $f_{\xi}(\nu)$. Hence, the light curves presented in \cite{sod05} can be used to derive the actual values of $B(\nu)$, $R(\nu)$ and $y(\nu)$.

X-ray emission was observed from SN\,2003L on day 40. Hence, focus will be on that date together with $\nu = \nu_{\rm max}$. From the light curves presented in \cite{sod05}, it is estimated that on this day, the light curve peaked for $\nu = 20$\,GHz and that the corresponding peak flux was 3.1\,mJy. As shown  in the Appendix, this results in  $\nu_{\rm max} = 30$\,GHz and  $F_{\rm max} = 2.8\times10^{28}$ for an assumed distance of 92\,Mpc. Fitting a power-law to the x-ray flux, \cite{sod05} found a luminosity in the 2-10\,keV band of $9.2\times10^{39}$. With the definition of luminosity used above, this number should be corrected for the x-ray band width, i.e., divided by a factor $\ln 5$ so that $L_{\rm x} = 5.7\times10^{39}$. Together with $L_{\rm bol} = 1.5\times 10^{42}$, equations (\ref{eq6}), (\ref{eq7}) and (\ref{eq8}) then lead to
\begin{equation}
	B_{\rm max} = 0.68\left(\frac{e_{\rm rel,max}}{e_{\rm rel,ave}}\right)^{-2/5},
	\label{eq17}
\end{equation}
\begin{equation}
	R_{\rm max,16} = 0.66\left(\frac{e_{\rm rel,max}}{e_{\rm rel,ave}}\right)^{-1/10},
	\label{eq18}
\end{equation}
and
\begin{equation}
	y_{\rm max} = 2.4\times10^2\left(\frac{e_{\rm rel,max}}{e_{\rm rel,ave}}\right)^{19/10}.
	\label{eq19}
\end{equation}
Furthermore, from equation (\ref{eq14}), one finds
\begin{equation}
	R_{\rm min,16} = 1.3\left(\frac{e_{\rm rel,min}}{e_{\rm rel,ave}}\right)^{-1/10},
	\label{eq20}
\end{equation}
which gives for the velocity of the forward shock $v_{\rm sh}\geq R_{\rm min}/t = 3.8\times10^9(e_{\rm rel,min}/e_{\rm ave})^{-1/10}$.

The main thing to notice from equations (\ref{eq17}) - (\ref{eq19}) is the large value for $y_{\rm max}$, unless the source is highly inhomogeneous (i.e., $e_{\rm rel,max} \ll e_{\rm rel,ave}$). For a homogeneous source, this implies both a small value for $B$ (see equation (\ref{eq1})) and a large value for $U_{\rm rel}/U_{\rm B}$ (see equation (\ref{eq3})). This is contrary to the conclusion drawn in \cite{sod05} that the observed x-ray emission is consistent with a homogeneous source as well as equipartition between relativistic electrons and magnetic fields. The claim is based on a lower limit for the value of $B$, which they derived from the apparent absence of cooling in the observed radio light curves. However, the limit they deduced is actually an upper limit to $B$ (i.e., a lower limit to $U_{\rm rel}/U_{\rm B}$). 

The need for a large value of $U_{\rm rel}/U_{\rm B}$ in a homogeneous source is not due to the assumption of an inverse Compton scattering origin of the x-ray emission. This can be seen from the calculations in \cite{c/f06}. They fitted a standard homogeneous model to the radio observations of SN\,2003L and found that the deduced magnetic field implied $\nu_{\rm cool,10} = 0.87\,t_{10} (U_{\rm rel}/U_{\rm B})^{12/19}$. The highest frequency for which \cite{sod05} present a light curve is 22.5\,GHz. It peaks at around 35\,days. Assuming equipartition between electrons and magnetic field gives $\nu_{\rm cool,10} = 3.0$; in fact, the deduced source parameters indicate that radiative cooling due to inverse Compton scattering cannot be neglected and, furthermore, they used $t_{\rm ad} = t$. With the inclusion of both inverse Compton scattering and a longer adiabatic time-scale, the value of $\nu_{\rm cool}$ decreases below that of $\nu_{\rm abs}$. Such a low value for $\nu_{\rm cool}$ would severely affect both the spectra and the time evolution of the peak flux of the light curves. Hence, the observed absence of radiative cooling indicates large values for $U_{\rm rel}/U_{\rm B}$ for a homogeneous model, irrespective of the origin of the x-ray emission.

The effects of the apparent absence of radiative cooling in an inhomogeneous model can be estimate from equation (\ref{eq10a}). The observed flattening of the spectra makes the optically thin emission to be dominated by $\nu_{\rm max}$. On day 40, one finds for this subcomponent $\nu_{\rm cool,10}^{\rm peak} = 1.0/\phi_{\rm los,max}^2 $, where $m = 0.9$ has been assumed. In their modelling of SN\,2003L, \cite{sod05} found no need to change the optically thin spectral index as the value of $\nu_{\rm abs}$ declined by a factor five. As an example, $\nu_{\rm cool,10} = 10^2$ on day 40 results in $F_{\rm max} \propto [1 + (\nu_{\rm max}/\nu_{\rm cool})^{1/2}]^{-5/7} = 0.89$ and $\Delta \alpha(2\nu_{\rm abs}) = 9.8\times 10^{-2}$. Since observations were consistent with $\alpha \approx 1.1$ , it will be assumed that changes in $\alpha$ by more than 0.1 would have been noted. Hence, this will be taken as a lower limit to $\nu_{\rm cool}$. This implies a suppression of $F_{\rm max}$ less than 11\,\% on that day due to radiative cooling, which is consistent with the roughly constant value $F_{\rm max}$ during the observed time period. This lower limit on the cooling frequency then implies $\phi_{\rm los,max} < 0.1$. Note that this upper limit is quite robust, since it does not depend on the actual value of $y$, i.e., whether cooling is due to synchrotron radiation or inverse Compton scattering.

For the sub-component with $\nu = \nu_{\rm max}$ at day 40, equation (\ref{eq10}) can be written as 
\begin{equation}
	\nu_{\rm cool,10} = 0.81\frac{y_{\rm max}^{12/19}}{\phi_{\rm los,max}^2\left(1 + 0.28y_{\rm max}^{10/19}\right)^2}.
	\label{eq21}
\end{equation}
Further constraints on the inhomogeneities can be obtained by assuming the x-ray emission to be due to inverse Compton scattering, since the value of $y_{\rm max}$ is then given by equation (\ref{eq19}). With $y_{\rm max} = 2.4\times 10^2\,(\phi_{\rm los,max} \delta_{\rm max})^{19/10}$, equation (\ref{eq21}) leads to
\begin{equation}
	\nu_{\rm cool,10} = \frac{26\,\delta_{\rm max}^{6/5}}{\phi_{\rm los, max}^{4/5}\left(1+5.0\,\phi_{\rm los, max} \delta_{\rm max}\right)^2}.
	\label{eq22}
\end{equation}
It is also seen that $U_{\rm ph}/U_{\rm B} = 5.0\,\phi_{\rm los,max} \delta_{\rm max} <  0.50\,\delta_{\rm max}$, so that $\delta_{\rm max} <  2.0$ implies that cooling is dominated by synchrotron radiation.

In order to proceed further, an assumption regarding the spatial distribution of relativistic electrons needs to be made. As an example, assume that the relativistic electrons are injected uniformly at a forward shock, while the inhomogeneities are due to the process amplifying the magnetic field in the inter-shock region. This corresponds to $\delta_{\rm max} = 1$. Together with $\nu_{\rm cool,10} = 10^2$, it is then found from equation (\ref{eq22}) that $\phi_{\rm los,max} = 7.9\times 10^{-2}$, which yields $y_{\rm max} = 1.9$.  Since, in this case, synchrotron cooling dominates over inverse Compton scattering, equation (\ref{eq22}) is valid irrespective of the location of the synchrotron emission region(s) within the source (see section \ref{sect2c}).

With the use of equation (\ref{eq3}), this value of $y_{\rm max}$ constrains the value of $\gamma_{\rm min}U_{\rm rel}/U_{\rm B}$ . It should be noted that it is not possible to deduce a value for $U_{\rm rel}$ alone. The reason is that for p\,=\,3, the combination $\gamma\,U_{\rm rel}(>\gamma)$ is independent of $\gamma$, where $U_{\rm rel}(>\gamma)$ is the energy density of electrons with Lorentz factors larger than $\gamma$. The expression for $y$ in equation (\ref{eq3}) can then be rewritten as
\begin{equation}
	y_{\rm max} = \frac{\gamma_{\rm min} U_{\rm rel}}{U_{\rm B}}\left(\frac{\phi_{\rm los,max}}{\phi_{\rm los,min}}\right)^{1/10}\phi_{\rm los,max}
	\label{eq23}
\end{equation}
where the definitions $R(\nu)/R_{\rm o} = \phi_{\rm cov}(\nu)^{1/2} $ and $\phi_{\rm los}(\nu) = 2 R_{||}(\nu)/R_{\rm o}$ have been used together with equation (\ref{eq14}). The value of $y_{\rm max}$ deduced above from observations and neglect of the weak dependence on the variations of $\phi_{\rm los}$ over the source lead to $\gamma_{\rm min}U_{\rm rel}/U_{\rm B} \approx 24$. This is an upper limit, since the value used for $\nu_{\rm cool}$ is a lower limit. 

The only limit on the value of $\gamma_{\rm min}$ comes from the observed absence of any cut-off in the electron spectrum; i.e., $\gamma_{\rm min} < \gamma_{\rm abs}$, where $\gamma_{\rm abs} = (\nu_{\rm abs}/\nu_{\rm B})^{1/2}$ \cite[where the numerical coefficient is appropriate for optically thick synchrotron radiation and p\,=\,3,][]{bjo22}. 
With the value of $y_{\rm max}$ derived from observations, equation (\ref{eq1}) gives $\gamma_{\rm abs} = 76$. Hence, it is not possible to argue for a value of $U_{\rm rel}/U_{\rm B}$ larger than unity. Since $\gamma_{\rm min}U_{\rm rel}$ is assumed constant, one may note from equation (\ref{eq15}) that the value of $U_{\rm rel}/U_{\rm B}$ varies by a factor $8.3(\phi_{\rm los,min}/\phi_{\rm los,max})^{4/5}$ from $\nu = \nu_{\rm min}$ to $\nu = \nu_{\rm max}$. This can be a rather large factor and, for example, the sub-component giving rise to the observed optically thin synchrotron flux may have rough equipartition between relativistic electrons and magnetic field, while for most of the source the electrons are dominating the energy density.

\subsection{SN\,2002ap}\label{sect3b}
Radio observations of SN\,2002ap was made by \cite{ber02}. Optically depth effects was observed at 1.43\,GHz only and, in addition, the observations were affected by interstellar scattering and scintillation. They modelled the time evolution of the radio flux by making a multi-parameter fit to the light curves from which the most likely solution was determined. One of their main conclusions was that the optically thin light curves implied p\,=\,2. However, the optically thin spectral index was $\alpha \approx 0.9$, indicating instead p $\approx$ 2.8. 

This discrepancy led \cite{b/f04} to suggest that the observed steepening of the spectrum was caused by radiative cooling. The effects of radiative cooling were determined with the use of the standard homogeneous model. Also, x-ray emission was observed on day 6 \citep{s/k02,sut03} and it was assumed that this corresponded to the inverse Compton scattered radiation. As a result, for this day, a unique solution for $R$, $B$, and $y$ can be obtained and, hence, the cooling frequency can be predicted. Although a steepening of the spectra due to radiative cooling is clearly seen in figure\,1 in \cite{b/f04}, the cooling frequency is too large to account for the spectral index in the observed frequency range. One should note though that $t_{\rm ad} = t$ was used in the calculations. As discussed in section\,\ref{sect2c}, the actual value for $t_{\rm ad}$ is a factor $3/2{\rm m}$ larger. This will decrease the value of $\nu_{\rm cool}$ (see equation \ref{eq10}); however, since p\,=\,2.1 was used in the calculation, this does not steepen the spectra enough to be consistent with observations.  

When radiative cooling is not important, the light curve for the optically thin synchrotron radiation varies with time as $F_{\nu}  \propto v_{\rm sh}^3 t^3 B^{(p+5)/2}$, where $U_{\rm rel} \propto U_{\rm B}$ and $\gamma_{\rm min} =$\,constant have been assumed. It is common to assume that the energy density of the magnetic field scales with the thermal energy density behind the shock so that $B \propto t^{-1}$. With $p\,=\,2$, this gives $F_{\nu}  \propto v_{\rm sh}^3 t^{-1/2}$. Since $v_{\rm sh} \propto t^{-0.12}$ is thought appropriate for type 1b/c supernovae \citep{m/m99}, one finds $F_{\nu} \propto t^{-0.86}$, which is close to the observed relation ($F_{\nu}^{\rm obs} \propto t^{-0.8}$). However, in order to account for the optically thin spectral index, a value for the cooling frequency is then required, which lies below the observed range. However, such a value is unlikely for two reasons: (1) As already mentioned, it is hard to make it compatible with a homogeneous source. (2) \cite{b/f04} showed that $t_{\rm cool}/t$ reaches a minimum around 10\,days. Hence, a dip in the light curves is expected around this time. As a consequence, the column density of relativistic electrons would decrease with time slower than in the non-cooling case after this date, which would lead to a flattening of the light curves. Hence, in order for the standard model to be compatible with observations, a value of p larger than 2 is required.

The properties of type Ib/c supernovae have been discussed by \cite{c/f06}. They showed that both spectra and light curves are consistent with p\,$\approx$\,3 and a rather small spread of p-values. Although spectra as well as light curves for SN\,2002ap are somewhat flatter than the average, they do not stand out as being qualitatively different. As mentioned above, if the source structure in SN\,2002ap is homogeneous, radiative cooling is indicated. This would increases the difference in p-values even further between SN\,2002ap and  the rest of the type Ib/c supernovae, since, for them, radiative cooling is unlikely to affect the spectrum. 

Hence, there are two scenarios for SN\,2002ap: A homogeneous source structure implies an unusually flat energy distribution for the injected relativistic electrons, while an inhomogeneous structure would allow properties more in line with the rest of the type Ib/c supernovae.

 In order for SN\,2002ap to remain a typical Ib/c supernova, the cooling frequency needs to be larger than the value deduced for a homogeneous source. One may notice that the peak of the light curve at 1.43\,GHz in SN\,2002ap is broader than expected for a homogeneous source. It is hard to judge whether this is intrinsic to the source or due to the effects of interstellar scattering and scintillation. An intrinsic origin would imply an inhomogeneous source structure. Hence, there are indications that also the emission region in SN\,2002ap is inhomogeneous. 

 It is estimated from \cite{ber02} that on day 6, $\nu_{\rm abs} = 3.0$\,GHz and $F_{\nu_{\rm abs}} = 0.40$\,mJy, which, for a distance of 7.3\,Mpc, corresponds to $F_{\nu_{\rm abs,27}} = 2.6\times10^{-2}$. Note that these values differ somewhat from those in \cite{ber02}, since their definition of $\nu_{\rm abs}$ corresponds to $\tau = 1$ rather than the spectral peak. Furthermore, $L_{\rm x} =2.0\times10^{37}$ \citep{sut03}, where a band width of 0.3 - 10\,keV has been used to convert the observed x-ray emission to $L_{\rm x}$, and $L_{\rm bol} = 1.6\times 10^{42}$ \citep{pan03}. 
 
This yields from equations (\ref{eq10a}), (\ref{eq10b}) and (\ref{eq8}), $y_{\rm cool,10}^{\rm peak} = 9.3/\phi_{\rm los}^2$, $y_{\rm peak} = 1.2 \times 10^{-3}$ and $y = 1.2 \times 10 (\delta\,\phi_{\rm los})^{19/10}$. Hence, $\hat{y} \equiv y/y_{\rm peak} = 1.0\times 10^4 (\delta\,\phi_{\rm los})^{19/10}$. In order for synchrotron radiation to dominate the cooling $\hat{y} < 1$ is needed (see figure\,\ref{fig1}), which requires $\delta\,\phi_{\rm los} <  7.9\times 10^{-3}$. Although such a low value cannot be excluded, it implies a source that is strongly magnetically dominated (see equation (\ref{eq3})). Hence, it is likely that inverse Compton scattering dominates the cooling, which gives $\nu_{\rm cool,10} = 0.54/(\delta^{4/5}\,\phi_{\rm los}^{14/5})$. It is seen that the cooling frequency in this case is quite sensitive to inhomogeneities; in particular, the value of $\phi_{\rm los}$. As a result, even a rather small deviation from homogeneity would increase the value of $\nu_{\rm cool}$  enough to make SN\,2002ap similar to other type Ib/c supernovae. The major difference between SN\,2002ap and SN\,2003L would then be that the latter is substantially more inhomogeneous. A similar conclusion was reach in \cite{bjo13} based on the observed x-ray emission.
 
The source parameters for SN\,2002ap can be calculated from equations (\ref{eq6}), (\ref{eq7}) and (\ref{eq8}) as
\begin{equation}
	B(\nu)=0.27 \left(\frac{e_{\rm rel}(\nu)}{e_{\rm rel,ave}}\right)^{-2/5},
	\label{eq24}
\end{equation}
\begin{equation}
	R_{16}(\nu) = 0.28 \left(\frac{e_{\rm rel}(\nu)}{e_{\rm rel,ave}}\right)^{-1/10}
	\label{eq25}
\end{equation}
and
\begin{equation}
	y(\nu)= 1.2\times 10 \left(\frac{e_{\rm rel}(\nu)}{e_{\rm rel,ave}}\right)^{19/10}.
	\label{eq26}
\end{equation}
This results in a velocity of the forward shock  $v_{\rm sh}\geq R_{\rm min}/t = 5.4\times10^9(e_{\rm rel,min}/e_{\rm ave})^{-1/10}$. Furthermore, with only small deviations from homogeneity, equation (\ref{eq26}) implies $y\,\sim\,12$, which suggests a value for $\gamma_{\rm min}U_{\rm rel}/U_{\rm B}$ similar to the one for SN\,2003L. Therefore, it is possible that the physical properties in SN\,2002ap are quite similar to those in the sub-component corresponding to $\nu = \nu_{\rm max}$ in SN\,2003L; in particular, both of them are consistent with having equipartition between relativistic electrons and magnetic fields. 

\section{Discussion}\label{sect4}
It has been clear for some time that the standard model cannot account for the observations of all radio supernovae. In most cases, this is due to wider/flatter spectra and light curves around their peak-values than predicted by synchrotron self-absorption. In order to accommodate the observations,  the standard model has been used as a starting point but then various modifications have been introduced; for example, \cite{sod05} changed the synchrotron spectral emissivity and acceptable $\chi^2$-values for the modelling of SN\,2011dh could only be achieved by artificially increasing the measured errors by factors 3 - 7  \citep{sod12,kra12}. 

\subsection{Inferred parameters for an inhomogeneous source}\label{sect4a}

\cite{b/k17} suggested that a physically realistic modification would be to assume that the deviations to the standard model were due various amounts of inhomogeneities. They gave a qualitative discussion of the characteristics of the inhomogeneities that could be deduced from the observations; in particular, it was emphasized that, roughly, there are two types of inhomogeneities that can be distinguished observationally. With a power-law distribution of electron energies, variations of the magnetic field strength along a given line of sight does only marginally affect the spectral distribution of the emitted radiation. Hence, small-scale inhomogeneities that uniformly fill the source give rise to an integrated spectrum similar to that predicted by the standard model. The main difference would be a reduced line of sight extension of the source, i.e., a smaller $\phi_{\rm los}$-value and, for example, the magnetic field strength deduced from the observations would correspond to its emission averaged value.

On the other hand, large scale inhomogeneities would cause the optical depth (i.e., the self-absorption frequency) to vary over the source and result in a flattening of the spectrum around its peak. This then leads to a covering factor ($\phi_{\rm cov}$), which varies with frequency in the flattened part of the spectrum. It was shown in section\,\ref{sect2b}, how the standard model can be reformulated to include such inhomogeneities and, hence, make it possible to use the observations to quantify the inhomogeneities in terms of $\phi_{\rm cov}$ and $\phi_{\rm los}$.

The standard model has three free parameters. When the inverse Compton scattered radiation is observed, in addition to the synchrotron self-absorption, they can be uniquely determined. Alternatively,  such a closure relation can be obtained also by a direct measurement of the outer radius ($\equiv R_{\rm VLBI}$) through spatially resolved VLBI-observations. When both are observed, it is possible to test for the presence of inhomogeneities even if spectra/light curves show no apparent signs of flattening. 

As discussed in section\,\ref{sect2c}, another possibility to constrain deviations from homogeneity is given by radiative cooling. For supernovae, this can be the case even if only a lower limit to the cooling frequency is obtained. The reason is that synchrotron radiation dominates the cooling for small values of $U_{\rm rel}/U_{\rm B}$, while inverse Compton scattering dominates at large values. Hence, there is a maximum value for the cooling frequency. If the observations indicate a cooling frequency larger than this maximum, inhomogeneities are implied. Such a conclusion is independent of the observed flux of inverse Compton scattered radiation. It was shown in section\,\ref{sect3a} that this suggests a rather stringent upper limit for the value of $\phi_{\rm los}$ in SN\,2003L.

The broadening of the synchrotron self-absorption peak in SN\,2003L can be used to constrain the value of $\phi_{\rm cov}$ for the sub-component responsible for most of the observed optically thin radio emission. Together with the upper limit of $\phi_{\rm los}$, it was argued that its volume filling factor was at most a few percent. This component was consistent with having equipartition between relativistic electrons and magnetic fields ($U_{\rm rel}/U_{\rm B} \sim 1$). On the other hand, the indicated range of B-values was large enough for the electrons to dominate the energy density for most of the source. 

The observations of SN\,2002ap have been fitted using the standard model \citep{b/f04}. Although the measurement errors are rather large, it was argued in section\,\ref{sect3b} that there are several implications from such a model that instead point to an inhomogeneous structure. However, in this case, the volume filling factor would be substantially larger than deduced for SN\,2003L. The rough equipartition between relativistic electrons and magnetic fields deduced from a homogeneous model would then apply only for the dominant component.

The conclusion that the main components in SN\,2003L and SN\,2002ap are both consistent with equipartition between relativistic electrons and magnetic fields depends on the actual value of the relative density of relativistic electrons ($\delta$). It is seen from equation (\ref{eq3}) that $y= \gamma_{\rm min}(U_{\rm rel}/U_{\rm B})\phi_{\rm los} \phi_{\rm cov}^{-1/2}$. Together with equations (\ref{eq7}) and (\ref{eq8}), one then finds $\gamma_{\rm min}U_{\rm rel}/U_{\rm B}\propto \phi_{\rm los}^{4/5} \delta^{9/5}$. Furthermore, the values  $\phi_{\rm los}$ and $\delta$ are degenerate (see, for example, equation (\ref{eq22})). Hence, the deduced partition of energy between relativistic electrons and magnetic fields in the main synchrotron component is rather sensitive to the actual spatial distribution of relativistic electrons within the source.

Independent of the detailed source structure, the volume filling factor of the main synchrotron component in SN\,2003L is likely to be much smaller than in SN\,2002ap. It is also notable that the light curves peaked much later in SN\,2003L than in SN\,2002ap; for example, at 150 days versus 3 days for $\nu = 4.9$\,GHz, where the calculated light curves in \cite{b/f04} have been used
for SN\,2002ap. If these characteristics are related, it would suggest either that all sources have a large filling factor early on but that it declines with time or that the filling factor is constant but those peaking early have large ones.
 
 The only type Ib/c supernovae observed within a large enough time span to constrain the evolution of the filling factor is SN\,2003bg. \cite{sod06} have made extensive radio observations of this supernova together with x-ray observations at days 30 and 120. The value of $L_{\rm x}/L_{\rm bol}\propto U_{\rm rel,ave}R_{\rm o}$ increased by a factor 1.6 between these two epochs. This lead \cite{c/f06} to argue that the x-ray emission is unlikely to be due to inverse Compton scattering, since a decrease by almost a factor four is expected in the standard model under the assumption of a constant mass-loss rate of the progenitor star. However, also the radio emission declined less rapidly than expected. As shown in \cite{sod06}, this is due to a distinct achromatic jump, roughly, by a factor two in the optically thin radio light curves starting around 120 days. This suggests an inhomogeneous source structure and, as such, it is possible to attribute the x-ray emission to inverse Compton scattering. If the large x-ray luminosity at 120 days is due to the same structural change that caused the jump in the radio, the jump-related increase of $U_{\rm rel,ave}$ would have been, approximately, a factor 6. Since the jump in the radio indicates a much smaller density increase, the volume filling factor of the radio emitting electrons ($\phi_{\rm fil}$) and/or the value of $\delta$ must have decreased. 

 Although the conclusions from SN\,2003bg indicate a decrease in the values of $\phi_{\rm fil}$ and/or $\delta$ associated with the flux increase, it is not clear how this can be used to further an understanding of the possible changes taking place in supernovae like SN\,2003L with smooth light curves. Another alternative is to use the correlation $L_{\rm radio} \propto (L_{\rm x}/L_{\rm bol})^2$ found for the type Ib/c supernovae discussed in \cite{bjo13}. It is valid over more than three orders of magnitude in $L_{\rm radio}$ with a small dispersion. Since the radio emission can be written $L_{\rm radio} \propto U_{\rm B}U_{\rm rel} \phi_{\rm fil}R_{\rm o}^3$, the observed radio light curves are consistent with the standard model and a constant value for $\phi_{\rm fil}$. However, the expected decrease of $L_{\rm x}/L_{\rm bol}\propto U_{\rm rel,ave}R_{\rm o} \propto v_{\rm sh}/\delta t$ is too rapid for the correlation found in \cite{bjo13} to apply to individual supernovae, unless $\delta\,\propsim\,t^{-1/2}$. Hence, the most straightforward interpretation is that this relation is due to constant values of $\phi_{\rm fil}$ and $\delta$ in individual sources but that their values are smaller for supernovae that peak later in the radio. This is also consistent with a lack of any observed changes in the light curves/spectra over a factor of five in frequency/time in SN\,2003L .
 
 It should be noted, though, that an alternative is possible. If the mass-loss rate of the progenitor star decreased with time just before the supernova explosion, the density of the circumstellar medium into which the forward shock moves would decrease less rapidly than $R_{\rm o}^{-2}$. In order to be consistent with the observed radio light curves, the value of $\phi_{\rm fil}$ would need to decrease with time. In order to rule out such a scenario as well as possible time variations of $\delta$, x-ray observations need to be done at two different times also for supernovae with smooth light curves.
 
 \begin{table}
\begin {center}
\centerline{Scheme for analysing an inhomogeneous synchrotron source}
\begin{tabular}{lp{10cm}}
\hline\hline\\
&
\end{tabular}
\centerline{Standard model}\\[0.5ex]
\begin{tabular}{lclc}
\toprule\\
Observation of $L_{\rm x}$ or $R_{\rm VLBI}$ & $\Longrightarrow$ & Closure relation for $B$, $R$ and $y$ &\\
& & (equations (\ref{eq1a}) - (\ref{eq4a}) or equations (\ref{eq1}) and (\ref{eq2})) &\\[2ex]
\hline\\
[2ex]
\end{tabular}
\centerline{Inhomogeneities}\\[0.5ex]
\begin{tabular}{lp{2cm}lp{4.7cm}}
\toprule\\
Type of observation & & Indications of inhomogeneities &\\[3ex]
\midrule\\
1) Spectral shape,  &  & $\nu_{\rm min} < F(\nu) < \nu_{\rm max}$, & flat-topped spectra \\[2ex]
2) Cooling frequency,  & & $\nu_{\rm cool} > \nu_{\rm cool}^{\rm peak}(\phi_{\rm los}=1)$,  & equations (\ref{eq10}) and (\ref{eq10a})\\[2ex]
3) $L_{\rm x}$ and $R_{\rm VLBI}$,  & & $R \neq R_{\rm VLBI}$ & section \ref{sect4a}\\[2ex]
\hline\\[2ex]
\end{tabular}
\end{center}
\end{table}

\begin{table}
\begin{center}
\centerline{Parameter values}
\begin{tabular}{lp{1cm}l}
\toprule\\
1a) $\nu_{\rm min}$ and $\nu_{\rm max}$ & $\Longrightarrow$ & Range of relative values for $B$, $R$ and $y$ over the source\\
& & (see, for example, equations (\ref{eq14}) - (\ref{eq16}))\\[1ex]
1b) $\nu_{\rm min}$, $\nu_{\rm max}$ and $L_{\rm x}$ & $\Longrightarrow$ & Range of absolute values for $B$, $R$ and $y$ over the source\\
& & (equations (\ref{eq6}) - (\ref{eq8})) \\[2ex]
2a) $\nu_{\rm cool} > \nu_{\rm cool}^{\rm peak}(\phi_{\rm los}=1)$ & $\Longrightarrow$ & Upper limit to $\phi_{\rm los}$\\[1ex]
2b) $\nu_{\rm cool} < \nu_{\rm cool}^{\rm peak}$ & $\Longrightarrow$ & Two values for $y$ (synchrotron or inverse Compton cooling)\\
& & (equation (\ref{eq10}) and figure\,\ref{fig1})\\[1ex]
2c)  $\nu_{\rm cool}$ and $L_{\rm x}$ & $\Longrightarrow$ & Constraints on $\phi_{\rm los}$ and $\delta$ \\
& & (see, for example,  equations (\ref{eq21}) and (\ref{eq22})) \\[2ex]
3a) $r_{\rm VLBI} = R(\nu)$ & $\Longrightarrow$ & Closure relation for the sub-components $B(\nu)$, $R(\nu)$ and $y(\nu)$\\
(Spatially resolved observations) & & (equations (\ref{eq1}) and (\ref{eq2}))\\[1ex]
3b) $r_{\rm VLBI} = R(\nu)$ and $L_{\rm x}$ & $\Longrightarrow$ & Variation of the column density of relativistic electrons \\
& & over the source, i.e., $\phi_{\rm los} \delta$ (equations (\ref{eq6}) - (\ref{eq8}))\\[2ex]
\bottomrule
\end{tabular}
\end{center}
\caption{Analysis of a synchrotron source when observations in addition to $F_{\nu_{\rm abs}}$ and $\nu_{\rm abs}$ are available. {\it Additional observed quantities:} A flat-topped spectrum - $F({\nu})$ for $\nu_{\rm min} < \nu <\nu_{\rm max}$; x-ray luminosity - $L_{\rm x}$; cooling frequency - $\nu_{\rm cool}$; spatially resolved VLBI-observations - outer radius ($R_{\rm VLBI}$) and effective radius ($r_{\rm VLBI}(\nu))$. {\it Deduced parameters for the homogeneous sub-components:} Analogous to the standard model - magnetic field ($B$), radius ($R$) and $y$; structural properties of the sub-components - covering factor ($\phi_{\rm cov}$), reduced line of sight extension ($\phi_{\rm los}$) and relative density of relativistic electrons ($\delta$). In flat-topped spectra, these parameter-values depend on frequency. It is sometimes convenient to use instead the volume filling factor $\phi_{\rm fil} \equiv \phi_{\rm cov} \times \phi_{\rm los}$ and the relative column density of relativistic electrons $e_{\rm rel}/e_{\rm rel,ave} \equiv \delta \times \phi_{\rm los}$.}
\end{table}

 \subsection{ A few general comments}\label{sect4b}
The complexity of an inhomogeneous source makes it clear that a detailed description of its properties is  not possible. Hence, it is important to introduce a model for the inhomogeneities that captures at least some of their main characteristics. It is argued that overlapping, homogeneous sub-components are a suitable starting point. In comparison to the standard model, it has three additional free parameters; namely, the covering factor ($\phi_{\rm cov}$), the reduced line of sight extension of the source ($\phi_{\rm los}$) and the relative density of relativistic electrons ($\delta$). These can then be used to define the volume filling factor as $\phi_{\rm fil} \equiv \phi_{\rm cov} \times \phi_{\rm los}$ and the relative column density of relativistic electrons $e_{\rm rel}/e_{\rm rel,ave} \equiv \delta \times \phi_{\rm los}$,
 
In addition to flat-topped spectra/light curves and radiative cooling, spatially resolved VLBI-observations have the potential to further constrain the properties of inhomogeneities in radio supernovae. Although this has not yet been achieved for type Ib/c supernovae, it has been done for two type IIb supernovae; namely, SN\,1993J and SN\,2011dh. The implications of these observations will be discussed in a forthcoming paper. The main advantages of such observations are: (1) An independent measurement of the outer radius gives directly the average brightness temperature of the source. In the standard model, this leads to a value for $U_{\rm rel}/U_{\rm B}$. One may then consider whether this value is consistent with other deduced properties of the source; if not, this suggests the presence of inhomogeneities. (2) In a spatially resolved source, an effective radius as function of frequency ($r_{\rm VLBI}(\nu)$) can be defined from the intensity variations. This can be used to break the degeneracy between the B-value and the column density of relativistic electrons along the line of sight in sources with flat-topped spectra/light curves \citep[see][for a more detailed discussion]{b/k17}.  An attempt has been made in Table\,1 to summarize how these different kinds of observations can be used to gain an understanding of the source structure. 

Neglecting the presence of inhomogeneities can lead to conclusions regarding vital aspects of the source, which differ substantially from the actual ones. In this paper, focus has been on the partition of energy between relativistic electrons and magnetic fields. Another parameter sensitive to inhomogeneities, is the radius, which determines the velocity of an unresolved source. Of particular interest here is its evolution with time, since it is directly related to the ejecta structure of the supernova \citep{bjo22}. Since, usually, the value of $F_{\nu_{\rm abs}}$ varies only slowly with time, it is the variation of $\nu_{\rm abs}$, which constrains the dynamic of the forward shock (see equation (\ref{eq2})). The value of $\nu_{\rm abs}$, in turn, is quite sensitive to $\phi_{\rm los}$ so that even small variations of its value can have significant effects on the inferred properties of the ejecta.

The deduced value of $U_{\rm rel}/U_{\rm B}$ is a good example of the dichotomy between a homogeneous and an inhomogeneous source; for example: When $L_{\rm x}$ is observed - a high value (homogeneity) versus a low value (inhomogeneity) or with spatially resolved VLBI-observations - a low value (homogeneity) versus high value (inhomogeneity). In principle, the three kinds of observations discussed above should allow to determine the three free parameters (i.e., provide a closure relation also for an inhomogeneous source). However, in practise, there are several limitations: (1) Even if flat-topped spectra/light curves are observed, only a lower limit to the actual source radius can be inferred. (2) A well determined value for the cooling frequency is usually hard to obtain; often only a lower limit is deduced. (3) Even in a spatially resolved source, the spatial scale of the inhomogeneities usually remains unknown (i.e., the radius of a homogeneous sub-component may be smaller than $r_{\rm VLBI}$). 

Therefore, additional constraints can be important; for example, one of the main arguments against a homogeneous model for SN\,1993J is that the high value for the magnetic field implied a kinetic energy of the ejecta more than an order of magnitude larger than even the most favourable models could provide \citep{bjo15}. In order to incorporate such limits and uncertainties in a more stringent way than done in the present paper, a forward modelling approach is then to be preferred \citep{bjo22}; for example, Monte Carlo calculations can give an estimate of a likely range of parameter-values consistent with observations.

\section{Conclusion}\label{sect5}
It is emphasized that one should be cautious about drawing strong conclusions regarding the source properties based on fitting the standard model to observations. This includes also cases where the standard model gives a good fit. 

It is shown how a simple extension of the standard model can be used to account for the observed effects of inhomogeneities. Although a detailed description of their properties is not possible, a quantitative comparison between different supernovae can be made. 

The cooling frequency depends on the ratio between the radiative and adiabatic cooling timescales. Inhomogeneities can decrease the effective value of the latter timescale and thus affect both spectra and light curves.  

The competition between synchrotron and inverse Compton cooling implies a maximum value for the cooling frequency.  This results in two independent sets of parameter-values corresponding to, respectively, synchrotron and inverse Compton cooling. In order to discriminate between the two, additional constraints are needed. 

The main results of the analysis are:

1) The volume filling factor of the synchrotron emission region in SN\,2003L is low; possibly as low as a few percent.

2) There are indications that also the source structure of SN\,2002ap is inhomogeneous. However, even so, the volume filling factor of its radio emission region is substantially larger than in SN\,2003L.

3) Observations suggest that the volume filling factor for an individual type Ib/c supernova remains roughly constant with time but that it is smaller for those, which peak later at radio frequencies.

4) Observations are consistent with energy equipartition between relativistic electrons and magnetic fields in the main synchrotron emitting region.
 
\newpage

\appendix

\begin{center}
\bf{Appendix}
\end{center}

\section{The use of $f_{\xi}(\nu)$ to deduce observed parameter values}\label{A}
The value of $\nu_{\rm min}$ can be estimated by fitting a homogeneous source to the low frequency part of $f_{\xi}(\nu)$ with the constraints that they should overlap in the optically thick part of the spectrum and have the same value where the homogeneous source peaks \citep[for a more detailed discussion see][]{b/k17}. This leads to
\begin{equation}
	\left[1-\exp(-\tau_{\rm min}^{\xi}(\nu))\right]^{1/\xi} = 1-\exp(-\tau_{\rm hom}).
	\label{eqA1}
\end{equation}
Likewise, $\nu_{\rm max}$ is obtained by requiring a homogeneous source to overlap $f_{\xi}(\nu)$ in the optically thin part of the spectrum and that they should have the same value where the homogeneous source peaks
\begin{equation}
	\frac{\left[1-\exp(-\tau_{\rm max}^{\xi}(\nu))\right]^{1/\xi}}{\tau_{\rm max}} = \frac{1-\exp(-\tau_{\rm hom})}{\tau_{\rm hom}}.
	\label{eqA2}
\end{equation}

With $\tau_{\rm hom} = 0.64$ ($p=3$), equations (\ref{eqA1}) and (\ref{eqA2}) can be solved to find that for $\xi = 0.5$, $\tau_{\rm min} = 1.4$ and $\tau_{\rm max} = 0.096$. Since $\tau \propto \nu^{-7/2}$, this gives for SN\,2003L, $\nu_{\rm max}/\nu_{\rm min} = 2.1$. Also, $F_{\rm max}/F_{\rm min} = f_{\xi}(\nu_{\rm max})/f_{\xi}(\nu_{\rm min}) = (\tau_{\rm max}/\tau_{\rm hom})(\tau_{\rm min}/\tau_{\rm max})^{5/7} = 0.99$. Hence, the inhomogeneities in SN\,2003L are described by
\begin{equation}
	\phi_{\rm cov,max} = 0.25 \left[\frac{e_{\rm rel,max}}{e_{\rm rel,min}}\right]^{-1/5},
	\label{eqA3}
\end{equation}
\begin{equation}
	\frac{B_{\rm max}}{B_{\rm min}} = 2.9 \left[\frac{e_{\rm rel,max}}{e_{\rm rel,min}}\right]^{-2/5},
	\label{eqA4}
\end{equation}
and
\begin{equation}
	\frac{y_{\rm max}}{y_{\rm min}} = 0.24\left[\frac{e_{\rm rel,max}}{e_{\rm rel,min}}\right]^{19/10}.
	\label{eqA5}
\end{equation}

A value for $F_{\rm max}$ can be deduced from the peak of the light curve in a similar manner. As shown in \cite{b/k17}, the peak frequency in $f_{\xi}(\nu)$ corresponds to an optical depth $\tau_{\rm p} = \tau_{\rm hom}^{1/\xi}$ or $\tau_{\rm p} = 0.41$ ($\xi = 0.5$), which leads to $\nu_{\rm max}/\nu_{\rm p} = 1.5$. Furthermore, $F_{\rm max} = [f_{\xi}(\nu_{\rm max})/f_{\xi}(\nu_{\rm p})]F({\nu_{\rm p}})$ and $f_{\xi}(\nu_{\rm max})/f_{\xi}(\nu_{\rm p}) = 0.89$ ($\xi = 0.5$). With an observed flux of 3.1\,mJy at  $\nu_{\rm p} = 20$\,GHz, one finds $F_{\rm max} = 2.8\times10^{28}$ for an assumed distance of 92\,Mpc.

\end{document}